\documentclass{emulateapj}
\usepackage{apjfonts}
\lefthead{HWANG AND HAN} 
\righthead{MICROLENSING CHARACTERIZATION OF SPOTS}

\begin{document}
\title{Characterization of Stellar Spots in Next-Generation 
Microlensing Surveys}
\author{Kyu-Ha Hwang and Cheongho Han\altaffilmark{1}}
\affil{Department of Physics, 
Chungbuk National University, Cheongju 361-763, \\
Republic of Korea; kyuha,cheongho@astroph.chungbuk.ac.kr}

\altaffiltext{1}{corresponding author}


\begin{abstract}
One of the important microlensing applications to stellar atmospheres
is the study of spots on stellar surface provided by the high resolution 
of caustic-crossing binary-lens events. In this paper, we investigate 
the characteristics of spot-induced perturbations in microlensing 
light curves and explain the physical background of the characteristics. 
We explore the variation of the spot-induced perturbations depending 
on various parameters characterizing the spot and investigate how well 
these parameters can be retrieved from observations in high-cadence 
future lensing surveys. From this, we find that although it would not 
be easy to precisely constrain the shape and the surface brightness 
contrast, the size and location of the spot on the stellar surface 
can be fairly well constrained from the analysis of lensing light 
curves.
\end{abstract}



\section{Introduction}

Over the past decade, microlensing has developed into a powerful 
tool to study stellar astrophysics, especially stellar atmospheres. 
See the review of \citet{gould01}.  Microlensing application to 
stellar atmosphere is possible for events where the source star 
crosses the caustic.  The caustic represents the source position 
at which the lensing magnification of a point source formally becomes 
infinity.  Then, as the caustic passes over the face of the star, 
different parts of the star are strongly magnified at different 
times, making it possible to probe detailed structures on the 
surface of the source star.  One of the stellar properties that 
can be investigated by the high resolution provided by 
caustic-crossing microlensing events is the limb darkening of 
stars \citep{witt95,bogdanov95,gaudi99,albrow99,albrow01, afonso00,
gould01,heyrovsky03,fields03,abe03,dominik04,kubas05,cassan06}.

Another application of microlensing to stellar atmosphere is 
the study of irregular surface structures such as spots.   
\citet{heyrovsky00} and \citet{hendry02} investigated the 
sensitivity to spots for single-lens events and found that 
the spot signal can be detected for events where the lens 
transits the face of the source star.  \citet{han00} and 
\citet{chang02} demonstrated the feasibility of spot detections 
from the observation of caustic-crossing binary-lens events.

Despite the feasibility of spot detections demonstrated by 
theoretical studies, there exists no firm detection of a spot 
signal for any of the microlensing events detected so far.  
This might be due to the small fraction of stars with spots, 
but it is more likely that the current lensing surveys are not 
sensitive enough to catch the signal.  For a caustic-crossing 
binary-lens event, the duration of a spot-induced signal is 
\begin{equation}
t_\bullet \sim \iota_\bullet {2\rho_\star\over \sin \phi} t_{\rm E},
\end{equation}
where $\iota_\bullet$ is the fraction of the spot size relative 
to the source size,   $\rho_\star$ is the source radius normalized 
by the Einstein radius $\theta_{\rm E}$, $t_{\rm E}$ is the Einstein 
time scale, and $\phi$ is the angle between the source trajectory 
and the caustic line.  Considering that $t_{\rm E}\sim 20$ days 
for a typical Galactic event, the time scale is 
\begin{equation}
t_\bullet \sim 0.12\ \left({\iota_\bullet \over 0.1}\right)
\left( {R_\star \over R_\odot}
\right)\ {\rm hrs}.
\end{equation}
Then, the spot-induced perturbation lasts only a few hours even 
for an event associated with a giant source star.  On the other 
hand, the observational cadence of the current lensing surveys 
is several times per night, which is far less than that required 
to detect the spot-induced perturbations.

However, the situation will be greatly different in future 
lensing surveys that will continuously survey wide fields at 
a high cadence using very large format imaging cameras.  The 
OGLE collaboration recently upgraded their camera to widen 
the field of view from $0.4\ {\rm deg}^2$ to $1.4\ {\rm deg}^2$.  
The MOA collaboration plans to upgrade the telescope to a field 
of view of $4\ {\rm deg}^2$ (T.\ Sumi 2009, private communication).  
The `Korea Microlensing Telescope Network (KMTNet)' is a survey 
exclusive for microlensing that plans to achieve $\sim$ 10 minute 
sample using a network of three 1.6 m telescopes each of which 
is equipped with a camera of a $4\ {\rm deg}^2$ field of view.  
With these upgraded and new instruments, the next-generation 
surveys will be able to resolve the short time-scale perturbations 
induced by spots.

In this paper, we further explore the feasibility of microlensing 
studies of stellar spots.  We investigate the characteristics of 
spot-induced perturbations in microlensing light curves and 
explain the physical background of the characteristics.  We 
also investigate the feasibility of constraining the physical 
parameters of spots from the observations in future lensing 
surveys.

\section{Spot-Induced Perturbations}

\subsection{Characteristics}

\begin{figure}[th]
\epsscale{1.1}
\plotone{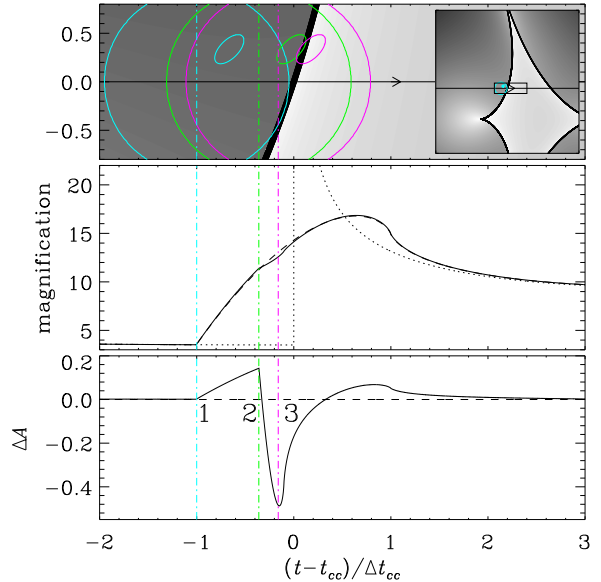}
\caption{\label{fig:one}
Light curve of a caustic-crossing binary-lens event occurred
on a source star with a spot (middle panel) and the the residual 
from that of the event occurring on an unspotted source star 
(lower panel).  The light curve is in lensing magnification 
and the residual is the difference in magnification.  The upper 
panel shows the geometry of the lens system where the thick 
linear structure represents the caustic and the straight line 
with an arrow represents the source trajectory.  The grey scale 
shows the point-source magnification as a function of position 
and brighter tone represents a higher magnification.  The inset 
shows the wider view of the geometry.  The circles with different 
colors represents the source at the times marked by numbers in 
the lower panel.  The spot of each source is also marked.  The 
three light curves in the middle panel represent those of events 
occurred on source stars with a spot (solid curve), without a 
spot (dashed curve), and a point source (dotted curve), respectively.  
The time is expressed with respect to the time of the caustic 
crossing of the source center, $t_{cc}$, in units of the 
caustic-crossing time scale $\Delta t_{cc}$.
}\end{figure}

Figure~\ref{fig:one} shows the light curve of an example 
caustic-crossing binary-lens event occurring on a source star 
with a spot and the resulting pattern of the spot-induced 
perturbation.  From the investigation of the perturbation, 
one finds the two following important characteristics of 
the perturbation pattern.  First, there exist both positive 
and negative deviations in lensing magnification from the 
light curve of the event occurring on an unspotted star.  
Second, the perturbation pattern is asymmetric with respect 
to the maximum deviation.

For an event occurred on a spotted star, the lensing 
magnification is represented by
\begin{equation}
A_\bullet={
\int_{\Sigma_\star} A(x,y) dxdy - (1-C_{\rm s}) \int_{\Sigma_\bullet} 
A(x,y) dxdy 
\over 
\Sigma_\star-\Sigma_\bullet
\label{eq3}
},
\end{equation}
where $A(x,y)$ represents the point-source magnification at 
a source position $(x,y)$, $C_{\rm s}$ is the flux ratio 
between the spot and unspotted regions, and $\Sigma_\star$ 
and $\Sigma_\bullet$ represent the surface areas of the star 
and the spot, respectively.  Here we assume that the brightness 
distribution on the source star surface other than the spot 
region is uniform.\footnote{For the effect of limb darkening 
on the spot-induced perturbation, see \citet{hendry02}} 
Then, the first term of the numerator of the right side of 
equation~(\ref{eq3}) represents the contribution of the 
unspotted region to the magnification (non-spot term), 
while the second term represents the contribution of the 
spot region (spot term).

The {\it negative} deviation in the spot-induced perturbation 
occurs when the magnification of the spot region is very high.
In this case, the decrease of the numerator in equation~(\ref{eq3}) 
by the spot term is more important than the decrease of the 
denominator by the spot area, resulting in a negative deviation.  
The magnification of the spot region is maximized when the spot 
is on the caustic.

The {\it positive} deviation occurs when the caustic is over 
the source surface but the spot is away from the caustic.  In 
this case, the magnification at the position of the spot is 
not high and the non-spot term in equation~(\ref{eq3}) dominates 
over the spot term.  Then, the decrease in the numerator is 
negligible while the decrease of the denominator remains the 
same regardless of the spot position.  With the increment of 
the denominator, the magnification is higher than the magnification 
of an unspotted event, resulting in a positive deviation.

The asymmetry of the perturbation pattern is caused by the 
difference in magnification patterns between the outside and 
the inside regions of the caustic.  In the outer region, the 
magnification plummets as the distance $d$ from the caustic 
increases, while the magnification inside the caustic decreases 
smoothly as $A\propto d^{-1/2}$.  Then, the spot term in 
equation~(\ref{eq3}) is important in a very narrow region 
outside of the caustic but in a much wider region inside the 
caustic.  As a result, the deviation pattern is asymmetric.

With the understanding of the physical background of 
the patterns of spot-induced perturbations, the locations 
of several important turning points in the perturbation can 
be explained.  For the case of the example event presented 
in Figure~\ref{fig:one}, there exist three such points, each 
of which is marked by a number.
\begin{enumerate}
\item
The position marked by ``1'' corresponds to the moment at 
which the source surface enters the caustic.  Differential 
magnification becomes important from this moment.  The highly 
magnified parts on the source star surface are unspotted 
regions and thus the non-spot terms dominates over the spot 
term, resulting in a positive deviation. 
\item
The position marked by ``2'' corresponds to the moment of 
spot's entrance into the caustic.  From this moment, the 
spot term becomes important and the deviation rapidly drops 
into negative values. 
\item
The negative deviation is maximized when the spot term is 
maximized and this corresponds to the position marked by 
``3''.  We note that the magnification inside and outside of 
the caustic is very different and thus the moment of maximum 
negative deviation occurs not at the moment when the center 
of the spot is exactly on the caustic but when the center is 
slightly shifted toward the inside of the caustic. 
\end{enumerate}

\begin{figure*}[th]
\epsscale{0.7}
\plotone{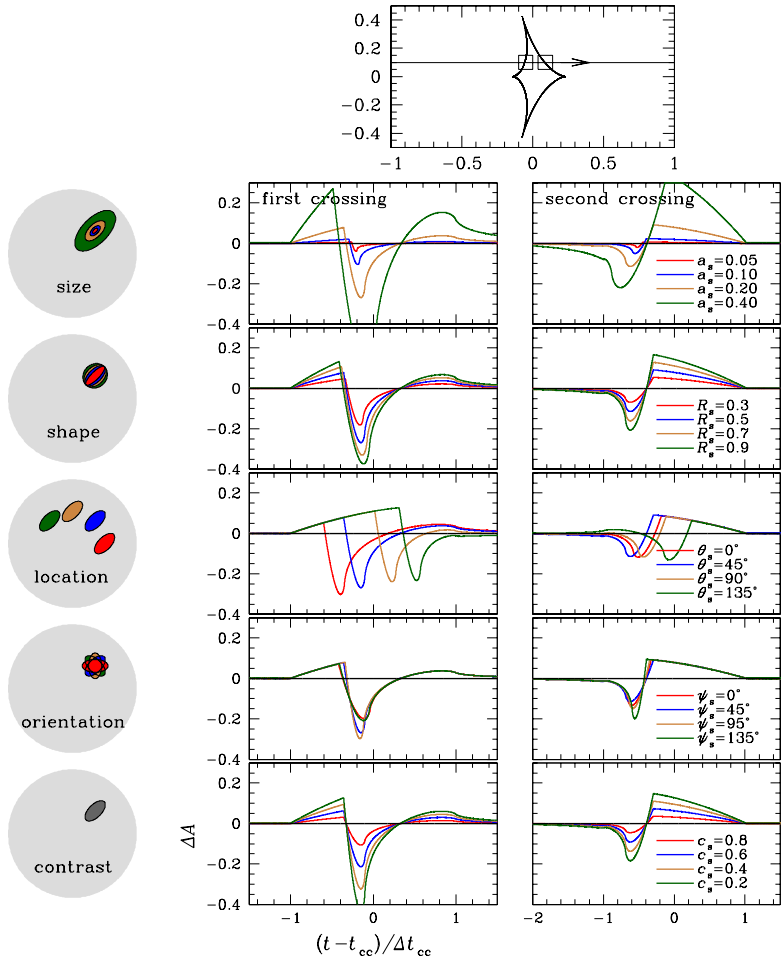}
\caption{\label{fig:two}
Variation of the spot-induced perturbations.  The top panel 
shows the geometry of the lens system with respect to the 
caustic produced by a binary lens system with a mass ratio 
$q=0.5$ and a projected separation normalized by the Einstein 
radius between the components of $s=0.7$.  The source radius 
relative to the Einstein radius is $\rho_\star=0.02$.  The 
subsequent panels show the variation depending on the spot 
size, the shape, the location, the orientation, and the 
contrast in brightness between the spot and unspotted regions,
respectively.
}\end{figure*}

\subsection{Variation}

Based on the fundamental scheme of the deviation described 
in the previous subsection, spot-induced perturbations take 
different shapes depending on various factors.  These factors 
include the size, shape, location on the source surface, 
orientation, brightness contrast with respect to the 
unspotted region, etc.  

Figure~\ref{fig:two} shows how the spot-induced perturbation 
depends on various factors.  Under the simplified 
approximation of a single elliptical spot with a uniform 
surface brightness, we parameterize the spot size as the 
semi-major axis $a_{\rm s}$ normalized by the source radius, 
the shape as the axis ratio $R_{\rm s}$ of the ellipse, the 
location on the source as the polar coordinates ($r_{\rm s}$,
$\theta_{\rm s}$) of the center of the ellipse with respect to 
the source center, the orientation as the angle $\psi_{\rm s}$ 
between the semi-major axis of the spot and the source trajectory, 
and the brightness contrast as the flux ratio $C_{\rm s}$ between 
the spot and unspotted regions.

\section{Characterization}

In the previous section, we explored the variation of the 
pattern of spot-induced perturbations depending on various 
parameters that characterize the spot.  In this section, we 
investigate how well these parameters can be retrieved from 
the analysis of light curves of event to be observed in future 
lensing surveys.  For this, we estimate the uncertainties of 
the spot parameters by fitting an example caustic-crossing 
binary-lens event produced by simulation under the observational 
conditions of planned future lensing surveys.

We choose the lensing parameters of the example event by 
adopting those of a typical Galactic bulge event occurred 
on a giant star.  The adopted values of the Einstein time 
scale is $t_{\rm E}=20$ days.  The projected separation 
normalized by the Einstein radius and the mass ratio between 
the lens components are $s=0.7$ and $q=0.5$, respectively.  
The apparent baseline magnitude of the source is $I=15$ and 
the fraction of the flux from the lensed star is 70\% 
and the rest comes from blended stars.  The source trajectory 
with respect to the caustic is same as shown in the top 
panel of Figure~\ref{fig:two}.  The values of the parameters 
characterizing the spot properties are $a_{\rm s}=0.3$, 
$R_{\rm s}=0.5$, $\psi_{\rm s}=45^\circ$, $r_{\rm s}=0.5$, 
$\theta_{\rm s}=45^\circ$, and $C_{\rm s}=0.5$.

\begin{figure}[th]
\epsscale{1.1}
\plotone{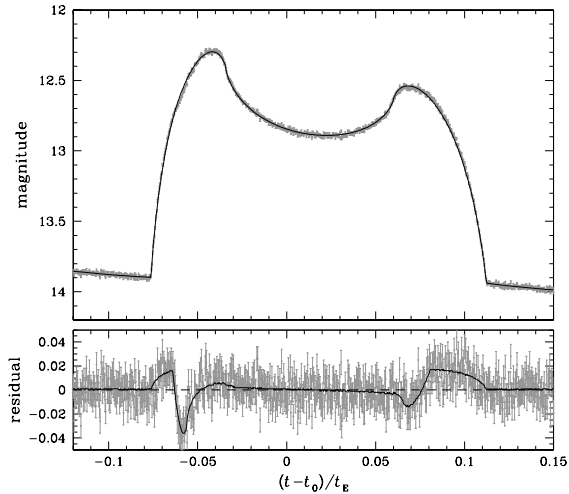}
\caption{\label{fig:three}
Light curve of the event produced by simulation that is used 
for the estimation of the uncertainties of the spot parameters.
We note that the light curve is expressed in magnitude.
The time $t_0$ represents the moment of the closest approach 
of the source star to the center of mass of the binary lens.
}\end{figure}

The event is assumed to be continuously observed in the $I$ 
passband by using a network of telescopes with a cadence of 
6 times per hour.  The exposure of each observation is 30 
seconds.  The aperture of each telescope is 1.6 m and the 
quantum efficiency of the detector is 0.8.  Given the 
specification of the instrument and the magnitude of the 
source and blend, we set the photometric uncertainty 
by assuming that the photometry follows photon statistics 
with a 1\% systematic uncertainty.  We also assume that the 
photometry is Gaussian distributed.  Figure~\ref{fig:three} 
shows the caustic-crossing part of the light curve produced 
by the simulation (upper panel) and the residual from the 
unspotted light curve (lower panel).

The uncertainties of the spot parameters are estimated by 
fitting the light curve produced by simulation.  In the 
fitting process, we search for a set of parameters describing 
the perturbation by minimizing $\chi^2$ in the parameter space.
We use a Markov Chain Monte Carlo method for the $\chi^2$ 
minimization.  Since a spot-induced perturbation occurs in a 
small region during the caustic crossing, its effect on the 
other lensing parameters is negligible.  We, therefore, set the 
lensing parameters fixed in the fitting process.\footnote{Another
reason for the use of this approximation is the limitation of 
computation time.  Modeling binary-lensing light curves requires 
to include many parameters. To describe spot-induced perturbations, 
at least 5 additional parameters are needed.  As a result, it 
is difficult to fir light curves letting all parameters vary.  
Fortunately, the  variation of the lensing parameters on the 
determined spot parameters will be very minor and thus will not 
affect the result of  the analysis.}
We hold two of the spot parameters fixed at a grid of values, 
while the remaining parameters are allowed to vary so that 
the model light curve results in a minimum $\chi^2$ at each 
grid point.  We set $a_{\rm s}$ and $R_{\rm s}$ as the grid 
parameters because they characterize the two most important 
properties of a spot.  Another important reason for using 
grid parameters is that it allows to use multiple CPUs by 
allocating computations of different sections of the 
grid-parameter space different CUPs.  Once a series 
of models for the individual sets of the grid parameters 
are obtained, we estimate the uncertainty of each parameter 
from the distribution of the parameter obtained by giving 
weight to the $\chi^2$ difference from the value of the 
best-fit model.

For the computation of lensing magnification affected by 
spot perturbation, we use the ray-shooting technique 
\citep{schneider86, kayser86, wambsganss97}.  In this 
technique, a large number of light rays are uniformly 
shot from the observer plane through the lens plane and 
then collected on the source plane.  We accelerate this 
process by   restricting the region of ray-shooting only 
in the region around the caustic and using a simple 
semi-analytic approximation in other parts of the source 
plane.  In addition, we keep the information of the 
positions of the light rays arriving at the target in 
the buffer memory of the computer so that it can be 
readily used for fast computation of the magnification.

\begin{deluxetable}{lcl}
\tablecaption{Uncertainties of Spot Parameters\label{table:one}}
\tablewidth{0pt}
\tablehead{
\colhead{property} &
\colhead{parameter} &
\colhead{uncertainty} 
}
\startdata
semi-major axis                 & $a_{\rm s}$                   & $\sigma_{a_{\rm s}}=0.04$                                        \\
axis ratio                      & $R_{\rm s}$                   & $\sigma_{R_{\rm s}}=0.2$                                         \\
orientation                     & $\psi_{\rm s}$                & $\sigma_{\psi_{\rm s}}=30^\circ$                                 \\
location of the source surface  & $(r_{\rm s},\theta_{\rm s})$  & $(\sigma_{r_{\rm s}},\sigma_{\theta_{\rm s}})=(0.02,5^\circ)$   \\
surface-brightness contrast     & $C_{\rm s}$                   & $\sigma_{C_{\rm s}}=0.4$            
\enddata
\end{deluxetable}

\begin{figure}[th]
\epsscale{1.1}
\plotone{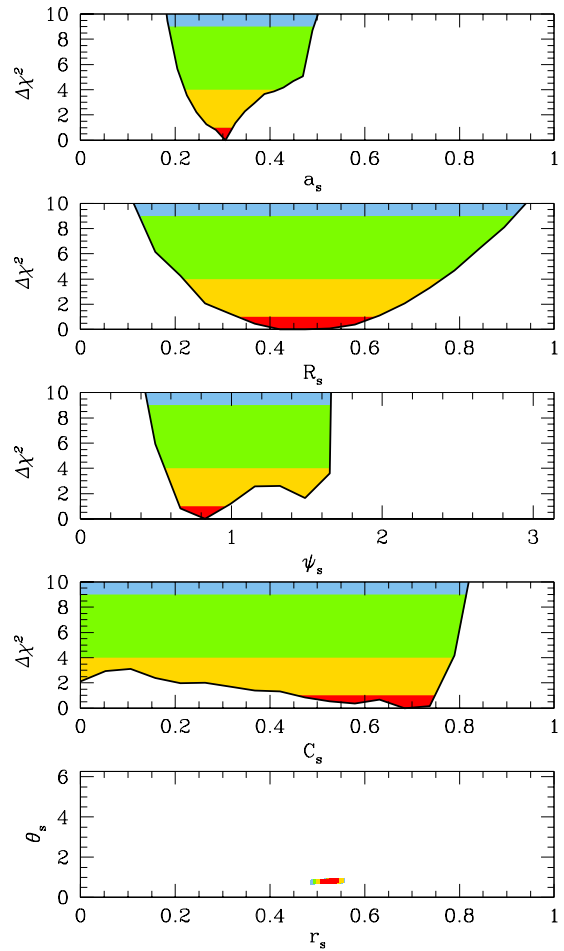}
\caption{\label{fig:four}
Distributions of $\Delta\chi^2$ for the spot parameters.
The regions marked in red, yellow, and green regions represent 
those of $\Delta\chi^2\leq 1$, 4, and 9, respectively.
}\end{figure}

We present the distributions of $\Delta\chi^2$ for the 
individual parameters in Figure~\ref{fig:four} and list 
the determined uncertainties in Table~\ref{table:one}.
From the uncertainties, we find the following factors.
First, the size of the spot can be well constrained with 
uncertainties of $\sigma_{a_{\rm s}}\sim 0.05$.  We 
judge the reason for this is due to the existence of the
discrete structures on the spot perturbation such as the 
turning points mentioned in section 2.  These structures 
depends on the spot size and thus help to constrain the 
size.  Second, the location of the spot on the surface of 
the source star is also well constrained.  This is because 
the source surface is swept by the caustic two times, 
i.e.\ at the entrance and exit.  For the individual caustic 
crossings, the incidence angles of the source respect to 
the caustic are different and thus the location of the 
spot is well constrained.  On the other hand, the shape 
of the spot is expected to be determined with a relatively 
large uncertainty.  Same is true for the surface-brightness 
contrast.  This can be understood from the similarity in 
the variation of the perturbation pattern depending on the 
two spot parameters of $R_{\rm s}$ and $C_{\rm s}$ as shown 
in Figure~\ref{fig:two}, where it is found that the variation 
of the perturbation with the increase of $R_{\rm s}$ is 
imitated by the variation with the increase of $C_{\rm s}$.
This implies that an observed perturbation suffers from the 
degeneracy between the two parameters and thus results in 
large uncertainties of both parameters.  The expected 
uncertainty of the orientation angle of the spot is 
$\sim 10^\circ$.

The uncertainties estimated in this work might be 
underestimated compared to the actual values in the future 
experiments.  Possible causes of the underestimation would 
be the simplification of the spot as a uniform ellipse and 
the disregard of detailed structures such as umbra and 
pen-umbra.  Considering that spots would occupy a small 
fraction of the source surface  and the estimated uncertainty 
of the spot-size is $\sim 5\%$ of the stellar size,  we think 
that the additional uncertainties caused by the disregard of 
very detailed structures would not be important.  Another 
cause would be the disregard of the stellar rotation.  Due 
to the stellar rotation, the locations of the spot on the 
stellar surface at the times of the caustic entrance and 
exit are different, causing additional uncertainties, 
especially in $r_{\rm s}$ and $\theta_{\rm s}$.  We note, 
however, that most of the events for which spots can be 
detected will be events associated with late-type giant 
source stars for which the angular rotation speed is small.  
Therefore, the uncertainty would not be seriously different 
from our estimation.

\section{Conclusion}

We investigated the characteristics of spot-induced perturbations 
in microlensing light curves and explained the physical background 
of the characteristics.  We also explored the variation of the 
spot-induced perturbations depending on various parameters 
characterizing the spot and investigated how well these parameters 
can be retrieved from observations in high-cadence future lensing 
surveys. From this, we found that the size and location of the 
spot on the stellar surface can be fairly well constrained from 
the analysis of lensing light curves, although it would not be 
easy to precisely constrain the shape and the surface brightness 
contrast

\acknowledgments 
This work is supported by Creative Research Initiative program
(2009-0081561) of National research Foundation of Korea.


\vskip2cm


\begin{thebibliography}{99}

\bibitem[Abe et al.(2003)]{abe03}
Abe, F., et al.\ 2003, \aap, 411, L493

\bibitem[Albrow et al.(1999)]{albrow99}
Albrow, M.\ D., et al.\ 1999, \apj, 522, 1011

\bibitem[Albrow et al.(2001)]{albrow01}
Albrow, M. D., et al.\ 2001, \apj, 549, 759

\bibitem[Afonso et al.(2000)]{afonso00}
Afonso, C., et al.\  2000, \apj, 532, 340

\bibitem[Bogdanov \& Cherepashchuk(1995)]{bogdanov95}
Bogdanov, M.~B., \& Cherepashchuk, A.~M.\ 1995, 
Astronomicheskij Zhurnal, 72, 873

\bibitem[Cassan et al.(2006)]{cassan06}
Cassan, A., et al.\ 2006, \aap, 460, 277

\bibitem[Chang \& Han(2002)]{chang02}
Chang, H.-Y., \& Han, C.\ 2002, \mnras, 335, 195

\bibitem[Dominik(2004)]{dominik04}
Dominik, M.\ 2004, \mnras, 353, 118

\bibitem[Fields et al.(2003)]{fields03}
Fields, D.\ L., et al.\ 2003, \apj, 596, 1305

\bibitem[Gaudi \& Gould(1999)]{gaudi99}
Gaudi, B.~S., \& Gould, A.\ 1999, \apj, 513, 619

\bibitem[Gould(2001)]{gould01}
Gould, A.\ 2001, \pasp, 113, 903

\bibitem[Han et al(2000)]{han00}
Han, C., Park, S.-H., Kim, H.-I., \& Chang, K.\ 2000, \mnras, 
316, 665

\bibitem[Hendry, Bryce \& Valls-Gabaud(2002)]{hendry02}
Hendry, M.\ A., Bryce, H.\ M., \& Valls-Gabaud, D.\ 2002, 
\mnras, 335, 539

\bibitem[Heyrovsk\'y(2003)]{heyrovsky03}
Heyrovsk\'y, D.\ 2003, \apj, 594, 464

\bibitem[Heyrovsk\'y \& Sasselov(2000)]{heyrovsky00}
Heyrovsk\'y, D., \& Sasselov, D. 2000, \apj, 529, 69

\bibitem[Kayser, Refsdal, \& Stabell(1986)]{kayser86}
Kayser, R., Refsdal, S., \& Stabell, R.\ 1986, \aap, 166, 36

\bibitem[Kubas et al.(2005)]{kubas05}
Kubas, D., et al.\ 2005, \aap, 435, 941 

\bibitem[Schneider \& Weiss(1986)]{schneider86}
Schneider, P., \& Weiss, A.\ 1986, \aap, 164, 237

\bibitem[Wambsganss(1997)]{wambsganss97}
Wambsganss, J.\ 1997, \mnras, 284, 172


\bibitem[Witt(1995)]{witt95}
Witt, H.~J.\  1995, \apj, 449, 42





\end{thebibliography}
\end{document}